\documentclass[aps,twocolumn,amssymb,superscriptaddress,nofootinbib,longbibliography]{revtex4-2}

\usepackage{graphicx}
\usepackage{color}
\usepackage{braket}
\usepackage{verbatim}
\usepackage{amsmath}
\usepackage{mathtools}
\usepackage{braket}
\usepackage{float}
\usepackage{hyperref}
\usepackage{dsfont}
\usepackage{lipsum}
\usepackage{tabularx,ragged2e}
\DeclareRobustCommand{\orderof}{\ensuremath{\mathcal{O}}}
\usepackage[ruled,vlined,linesnumbered]{algorithm2e}
\definecolor{darkblue}{RGB}{0,0,170}

\SetCommentSty{mycommfont}

\begin{document}
\title{Efficient percolation simulations for lossy photonic fusion networks}

\author{Matthias C. L\"{o}bl}
\email{matthias.loebl@nbi.ku.dk}
\affiliation{Center for Hybrid Quantum Networks (Hy-Q), The Niels Bohr Institute, University of Copenhagen, Blegdamsvej 17, DK-2100 Copenhagen {\O}, Denmark}

\author{Stefano Paesani}
\affiliation{Center for Hybrid Quantum Networks (Hy-Q), The Niels Bohr Institute, University of Copenhagen, Blegdamsvej 17, DK-2100 Copenhagen {\O}, Denmark}
\affiliation{NNF Quantum Computing Programme, Niels Bohr Institute, University of Copenhagen, Blegdamsvej 17, DK-2100 Copenhagen {\O}, Denmark.}

\author{Anders S. S\o{}rensen}
\affiliation{Center for Hybrid Quantum Networks (Hy-Q), The Niels Bohr Institute, University of Copenhagen, Blegdamsvej 17, DK-2100 Copenhagen {\O}, Denmark}

\begin{abstract}
The study of percolation phenomena has various applications ranging from social networks or materials science to quantum information. The most common percolation models are bond- or site-percolation for which the Newman-Ziff algorithm enables an efficient simulation. Here, we consider several non-standard percolation models that appear in the context of measurement-based photonic quantum computing with so-called graph states and fusion networks. The associated percolation thresholds determine the tolerance to photon loss in such systems and we develop modifications of the Newman-Ziff algorithm to perform the corresponding percolation simulation efficiently. We demonstrate our algorithms by using them to characterize exemplary fusion networks and graph states. The used source code is provided as an open-source repository. 
\end{abstract}

\maketitle

\section{Introduction}
In the field of quantum information, graph states are quantum states that can be represented by a graph: vertices are qubits, the fundamental storage unit of quantum information, and graph edges represent quantum mechanical entanglement between the qubits~\cite{Hein2006}. Such graph states are the key ingredient for measurement-based quantum computing~\cite{Raussendorf2001,Raussendorf2003}. A suitable large graph state (a so-called cluster state) provides a universal resource for this quantum computing paradigm~\cite{Raussendorf2001, Raussendorf2003} allowing quantum computation to be performed by doing measurements on the qubits. This approach is particularly suitable for photonic implementations of quantum information processing~\cite{GimenoSegovia2015, Paesani2022}. It is, however, challenging to realize a sufficiently large graph state of photons because, first, it is non-trivial to create edges/entanglement between photons in the graph state, and, second, photons can be lost with a non-negligible probability $p_{loss}$. Both of these limitations lead to imperfect graphs with some level of randomness, but tools from graph theory such as path-finding~\cite{Morley2017, Herr2018} and percolation models~\cite{Kieling2007, GimenoSegovia2015, Pant2019} can be employed to convert the imperfect graph state into a useful resource for quantum computing.

To create graph edges, probabilistic entangling operations~\cite{Browne2005, Knill2001} are typically considered for photonic qubits as deterministic photon-photon gates are difficult to realize~\cite{Hacker2016}. These can be implemented by joint two-qubit measurements called \textit{fusions} or \textit{Bell state measurements}, which consume the two measured photons (graph nodes). The desired measurement outcomes lead to successful fusions which, loosely speaking, make a connection between two graphs. Other measurement results herald fusion failure which does not generate the connection. Using fusions, many small graph states (named \textit{resource states}) can be connected to create larger entangled states~\cite{GimenoSegovia2015, Pant2019, Lobl2023b}. If a large entangled graph state is generated by such a process, it can be converted into a resource for measurement-based quantum computing through a process called lattice renormalization~\cite{Herr2018}. Whether applying many fusion operations leads to such a large graph state depends on the probability of fusion success, $p_s$, and the connectivity of all fusions. In the absence of loss, the required value for $p_s$ can correspond to a bond percolation threshold $\lambda_{bond}$ of the \textit{fusion network} determined by the geometric arrangement of the fusions~\cite{Kieling2007, GimenoSegovia2015, Pant2019}. If $p_s<\lambda_{bond}$, the resulting graph state consists of small pieces that are not useful for quantum computing. For $p_s>\lambda_{bond}$, a large connected graph state is created that percolates the fusion network, spanning from one of its ends to the other. In the latter case, the size of the state scales linearly with the system size providing the desired resource for measurement-based quantum computing.

The second issue, photon loss, is problematic for practical realizations since system efficiencies $\eta=1-p_{loss}$ are so far significantly below unity. This problem is due to absorption or leakage inside the employed photonic circuits and especially due to the employed photon sources~\cite{Chanana2022, Uppu2021, Tomm2021, Ding2023}. A loss leads to a mixed quantum state~\cite{Varnava2006} and a pure state can only be retained by removing the neighborhood of the lost qubit (the qubits that are entangled with the lost qubit) from the graph by measurements~\cite{GimenoSegovia2015, Lobl2023b} (see Fig.~\ref{fig:1}(a)).

In this paper, we consider probabilistic fusion networks in the presence of loss. In these fusion networks, small resource states are fused together to make a large entangled state. We show that the required value for $\eta$ corresponds to a percolation threshold $\lambda_{\eta}$ of new percolation models. The efficiency must be above the corresponding percolation threshold $\lambda_{\eta}$ to generate a graph state that is useful for measurement-based quantum computing. The percolation model corresponding to the process of photon loss differs from standard percolation models (like bond- or site-percolation~\cite{Stauffer2018} or a combination of both~\cite{Wattendorff2024}) as photon loss is more destructive than a missing site in a site-percolation model. Therefore, new algorithms are required to simulate the corresponding percolation models efficiently. We extend the algorithm developed by Newman and Ziff~\cite{Newman2000,NewmanZiff2001} for bond- and site-percolation such that it can be used to efficiently perform these new percolation simulations.

We consider several percolation models: first, a given graph state suffers photon loss (percolation model 1). Second, many \textit{star}-shaped resource states are arranged on a periodic fusion network. Fusions are applied between \textit{fusion photons} on the \textit{leaf} nodes of neighboring resource states to connect them to a larger graph. We consider both purely photonic resource states (percolation model 2) as well as resource states where the central qubit is the spin of a quantum emitter, e.g. atom~\cite{Thomas2022} or an equivalent solid-state system~\cite{Warburton2013,Lodahl2015,Atature2018,Shandilya2022,Montblanch2023,Simmons2023} (percolation model 2'). Finally, we consider repeat-until-success fusion networks where quantum emitters can generate new fusion photons conditioned on the failure of previous fusions (percolation model 3).

We have implemented all algorithms in C and the code is publicly available as an open-source repository~\cite{Lobl2023}. We recently used this source code to simulate various lattices of several dimensions and found that their loss tolerance can strongly differ~\cite{Lobl2023b}. Notably, photonic approaches are not restricted to a particular spatial arrangement of the qubits. This makes it possible to create lattices with various connection patterns and this can be used to increase the robustness to photon loss.

\section{Graph states}
\label{sec_graph}
Within quantum information science, graph states are represented by a graph $G=(V, E)$ corresponding to a quantum state where the vertices of the graph, $V$, are the qubits and the edges, $E$, are entangling operations between the qubits. The graph state can be defined as
\begin{equation}
    \label{eq:graph}
    \ket{G}=\prod_{(i,j)\in E} C_{ij}\ket{+}^{\otimes{V}},
\end{equation}
where $C_{ij}=\left(\ket{00}\bra{00}+\ket{01}\bra{01}+\ket{10}\bra{10}-\ket{11}\bra{11}\right)_{ij}$ is a controlled-$Z$ gate (controlled phase gate) between qubits $i$ and $j$, $\ket{+}=\frac{1}{\sqrt{2}}(\ket{0}+\ket{1})$ is an equal superposition of the two basis states $\ket{0}$ and $\ket{1}$, and $\ket{+}^{\otimes V}$ represents the product state where all qubits $V$ are in this state~\cite{Hein2006}. The order in which the controlled-$Z$ gates $C_{ij}$ are applied is irrelevant as all gates $C_{ij}$ commute. Therefore, a controlled-$Z$ gate can be represented by an undirected graph edge $e_{ij}=(i,j)$, and the graph state can be represented by the set of edges. Since the controlled-$Z$ operation is an entangling operation when applied to two qubits in the $\ket{+}$ state, the edges represent the entanglement structure of the state. Graph states are part of both the larger classes of stabilizer states~\cite{Aaronson2004} and the class of hyper-graph states~\cite{Rossi2013}. For more detailed information on graph states, we refer the reader to Ref.~\cite{Hein2006}.

Single-qubit measurements can be used to manipulate graph states~\cite{Hein2006} and use them for universal quantum computing~\cite{Raussendorf2001}. For this manuscript, it is only necessary to consider the effect of $Z$-basis measurements that project the measured qubit on the computational basis $\{\ket{0},\ket{1}\}$ (the $Z$-basis). To understand the effect on the graph state, we pick a qubit $A\in V$ and write Eq.~\eqref{eq:graph} as:
\begin{equation}
    \prod_{k\in N(A)} C_{Ak}\ket{+}_A\ket{\theta} = \frac{1}{\sqrt{2}}\left(\ket{0}_A\ket{\theta}+\ket{1}_AZ_{N(A)}\ket{\theta}\right),
\end{equation}
where $\ket{\theta}=\prod_{(i,j)\in E, i\neq A \neq j} C_{ij}\ket{+}^{\otimes{V\setminus A}}$ is the graph state corresponding to the induced subgraph where node $A$ has been removed from the graph $G$. $N(A)=\{i\in V\,|\,\exists \,(A,k)\in E\}$ represents all qubits in the neighborhood of $A$ with $Z_{N(A)}$ representing a Pauli-$Z$ gate ($Z=\ket{0}\bra{0}-\ket{1}\bra{1}$) on all these qubits. Assume one measures qubit $A$ in the $Z$-basis. Projecting on the first $Z$-eigenstate, $\ket{0}_A$, yields the state $\ket{\theta}$ and thus corresponds to removing qubit $A$ and its edges from the graph. Projecting on $\ket{1}_A$ leads to the state $Z_{N(A)}\ket{\theta}$ corresponding to qubit $A$ being removed and a Pauli-$Z$ gate applied to all its neighbors before the measurement~\cite{Hein2006}. Since such single qubit gates do not change the entanglement properties, the resulting state has the same computational power as one without the gates applied and can typically be dealt with by reinterpreting the outcomes of later measurements. For this article, the outcomes of the $Z$-basis measurements are thus inconsequential and will be ignored, although one will have to keep track of the outcome in an actual experimental implementation.

\begin{figure*}
\includegraphics[width=2.0\columnwidth]{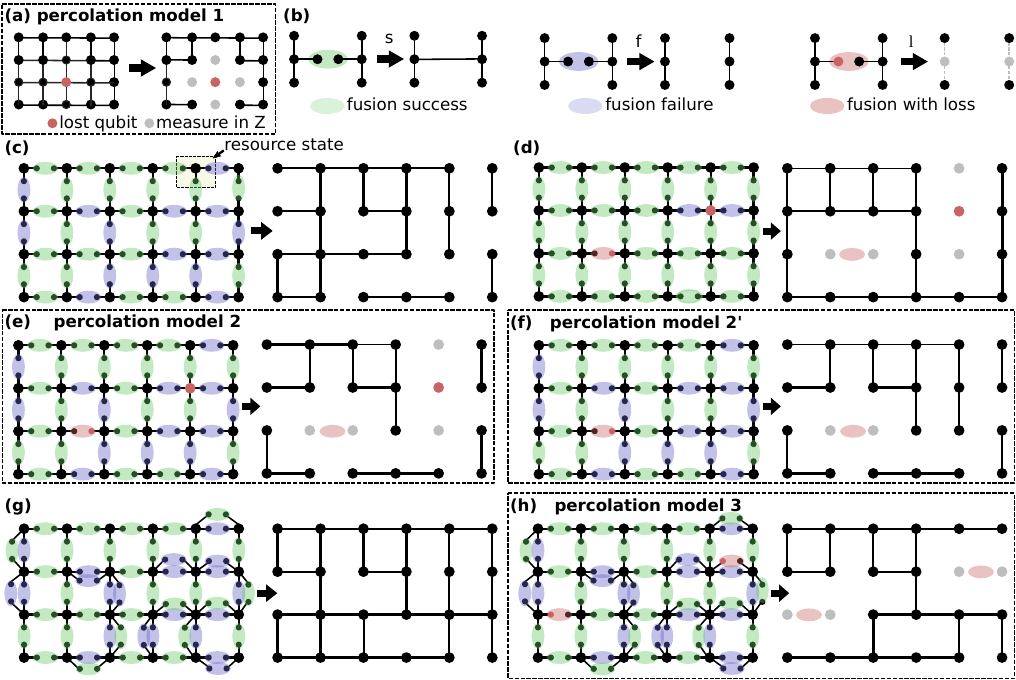}
\caption{\label{fig:1}The different percolation models. \textbf{(a)} Percolation model 1: loss on an existing graph state. Photon loss is compensated by measuring the neighborhood of a lost qubit in the $Z$-basis causing the deletion of its entire neighborhood. \textbf{(b)} Fusion of photons from two different graph states. The fusion can be realized by sending the two fusion photons (encircled by ellipses) to a suitable Bell state measurement setup~\cite{Browne2005, GimenoSegovia2015, Lobl2023b}. The two photons undergoing the Bell state measurement are removed from the graph, yet the measurement may lead to new connections being added to the graph depending on the measurement outcome. If the fusion succeeds (s), the neighbor of one fusion photon gets connected to the neighbor of the other one. If the fusion fails (f), there is no connection. If one or more fusion photons are lost, the neighborhoods of both qubits involved in the fusion need to be measured in the $Z$-basis, removing these nodes from the graph (gray). \textbf{(c)} Graph state generation from small resource states. Many star-shaped resource states are arranged on a square lattice and fused (for simplicity illustrated without photon loss). After the fusions, a new graph state is generated. To be of use for measurement-based quantum computing, the fusion success probability $p_s$ must be above the bond-percolation threshold of the fusion lattice. \textbf{(d)} Loss of a fusion photon/leaf qubit and loss of a central qubit. Loss of a fusion photon results in the deletion of the central qubit of two star-shaped resource states (left). If a central qubit of a star-shaped resource state is lost, all central qubits that are connected to it by a successful fusion need to be deleted (right). For this illustration, we have assumed that most fusions succeed. \textbf{(e)} Percolation model 2: photon losses and fusion failure happen at the same time. In this percolation model, there are losses of leaf and central qubits as illustrated. \textbf{(f)} Percolation model 2': only leaf qubits can be lost. \textbf{(g)} Example of a repeat-until-success scheme where failed fusions are repeated up to $n_{max}$ times by generating two new fusion photons. \textbf{(h)} Percolation model 3: Repeat-until-success scheme with photon loss.}
\end{figure*}

Loss of qubit $A$ qubit can be interpreted as measuring the qubit without knowing the outcome of the measurement, i.e. without knowing whether the gates $Z_{N(A)}$ have been applied to the remaining state $\ket{\theta}$ or not. The resulting quantum state is thus a mixed state. To retain a pure quantum state, the qubits in $N(A)$ need to be removed from the graph by measuring them in the $Z$-basis~\cite{GimenoSegovia2015} (see Fig.~\ref{fig:1}(a))~\footnote{This removes the same number $|N(A)|+1$ of qubits and so-called stabilizer generators~\cite{Aaronson2004} from the state. Instead, one could extract a maximum number of stabilizer generators without support on the lost qubit by Gaussian elimination and continue with the resulting mixed state. This would typically erase fewer stabilizer generators. If the considered lattice has no fault-tolerant properties, it is, however, unclear how to subsequently use the generated mixed state and we will not pursue this possibility any further.}.

\section{Definition of the percolation models}
In this section, we will define the percolation models, for which we will develop efficient algorithms.

\subsection{Percolation model 1: loss on graph states}
In the first model, we only consider the effect of loss on an already existing graph state (no fusions), where each qubit is represented by two states of a single photon, e.g. two different polarization states. Assume a given graph state $\ket{G}$ represented by a graph $G=(V, E)$. We only consider graphs that can be embedded as some periodic lattice but that is principally not necessary. Assume that $\ket{G}$ suffers photon loss with a rate of $p_{loss}=1-\eta$. As described before, the neighborhood of every lost photon needs to be measured in the $Z$-basis resulting in the neighbors being removed from the graph state. This corresponds to a modified site-percolation model where a missing site leads to a deletion of its entire neighborhood (see Fig.~\ref{fig:1}(a)). The corresponding percolation threshold $\lambda_{\eta}$ specifies the fraction of photons (graph nodes from $V$) that at least need to be present (not lost) such that in the limit of infinite lattice size there is a connected graph state left (some induced subgraph of $G$) which has a size that scales linearly with the size of $G$. Such a large graph state is required for measurement-based quantum computing and the loss tolerance of the graph state $\ket{G}$ is therefore characterized by $\lambda_{\eta}$ where $\eta=1-p_{loss}$ has to be above this number. Such a percolation model gives a guideline for the loss-tolerance of a given graph state, yet there are weaknesses regarding its practical applicability: first, photon loss is typically unknown (unheralded) before doing a destructive measurement~\cite{GimenoSegovia2015, Morley2017}, yet the required lattice renormalization (see section~\ref{sec:lattice_renormalize}) relies on knowing the losses before measuring the qubits in a certain basis. Without quantum non-demolition detection of whether a qubit is present or not~\cite{Stricker2020} it is therefore unclear to which degree the lossy graph state is useful for quantum computing eventually. Second, it is unclear how the large initial photonic graph state can be created physically.

\subsection{Percolation model 2, 2': loss on fusion network}
To alleviate the shortcomings of the model considered in the previous section we now switch to a model that includes the graph state generation. Fusion networks represent an approach to generate large graph states from smaller resource states that can more easily be generated experimentally~\footnote{Proposals for generating small resource graph states can be found in Refs.~\cite{Lindner2009, Buterakos2017, Tiurev2021} and recent experimental realizations in Refs.~\cite{Schwartz2016,Coste2022,Cogan2023,Thomas2022,Cao2023, Maring2023, Meng2023}.}. We consider star-shaped resource states and we assume that fusions~\cite{Browne2005} of photons at the leaf-nodes of the star-shaped resource states, the \textit{fusion photons}, are used to establish connections between the \textit{central} qubits of different star-shaped resource states\footnote{The considered star-shaped resource states are locally equivalent~\cite{Cabello2011,Adcock2020} to so-called GHZ states.}. Such a fusion network is illustrated in Fig.~\ref{fig:1}(c). As all the measurements associated with the fusions commute and the fusions performed are fixed (no adaptiveness), all fusions can be performed at once (ballistically~\cite{GimenoSegovia2015}). We only consider fusion networks where the resource states and the fusions can be arranged on a graph $G$ with a periodic embedding, the \textit{fusion lattice}. A node of $G$ represents a central qubit of a resource state and an edge $e_{ij}$ represents a fusion of two leaf nodes connected to the central qubits $i$ and $j$. The exemplary fusion network in Fig.~\ref{fig:1}(c) corresponds to a fusion lattice that is a square lattice, but arbitrary other lattice geometries are possible.

Fusions can be implemented using linear optics elements and photon detectors, with the measured photon pattern on the detectors heralding the successful implementation of the fusion~\cite{Browne2005,Gimeno2016}. Fusions are probabilistic, succeeding with a probability of $p_s$. We assume so-called rotated type-II fusion that consumes two fusion photons and connects the two central nodes of two star-shaped resource states upon success~\cite{GimenoSegovia2015,Gimeno2016,Lobl2023b}. The fusion fails with probability $1-p_s$ in which case the resource states are left mutually unconnected (see Fig.~\ref{fig:1}(b))~\footnote{A more detailed explanation of the required rotated type-II fusion can be found in Fig.~4 of Ref.~\cite{GimenoSegovia2015} or Fig. A5(d) of Ref.~\cite{Lobl2023b}.}. In the absence of photon loss, a large graph state is therefore generated once $p_s$ exceeds the bond percolation threshold of the fusion lattice (see Fig.~\ref{fig:1}(c))~\cite{Kieling2007, GimenoSegovia2015, Pant2019}. In the percolation models that we will consider in this section, $p_s$ can in principle be an arbitrary probability yet only certain values correspond to physical Bell measurement setups and we assume $p_s=0.5$ corresponding to the most simple setup\footnote{The fusion success probability can be improved (boosted) above $p_s=0.5$ by using additional photonic resources~\cite{Grice2011,Ewert2014,Witthaut2012} yet $p_s=0.5$ is easiest to implement in practice.}

Further, we assume that every photon, independent of whether it is used in a fusion or not, has a uniform probability $p_{loss}=1-\eta$ to be lost\footnote{A non-uniform distribution of loss probabilities may originate from different efficiencies of photon sources~\cite{Uppu2021, Tomm2021, Ding2023} or different paths that the photons take, but an investigation of this is beyond the scope of this work.}. When and where the photons are lost is generally not important as a loss will only be registered by the absence of a detection event in a measurement such as a fusion. Thus, the effect on the graph state only depends on the loss probability regardless of where the loss is applied. When one or more photons are lost in a fusion operation, a pure state is retained by measuring the neighborhood of both fusion photons (see Fig.~\ref{fig:1}(b))~\cite{GimenoSegovia2015, Lobl2023b}. The left part of Fig.~\ref{fig:1}(d) illustrates the effect of a lost fusion photon on a fusion lattice in the boundary case $p_s=1$. Losses in fusions are heralded~\cite{GimenoSegovia2015, Gimeno2016}, meaning that one knows about all lost fusion photons from the measured detection pattern, i.e. one knows that a photon is lost when it was not detected in the fusion process. A fusion with two photons attached to two central qubits $i$ and $j$ has three outcomes (see Fig.~\ref{fig:1}(b)): the fusion succeeds creating a connection between $i$ and $j$ (probability $p_s\cdot\eta^2$), the fusion fails to create a connection (probability $(1-p_s)\cdot\eta^2$), or a photon loss is detected (probability $1-\eta^2$) in which case $i,j$ need to be removed by $Z$-basis measurements (see section~\ref{sec_graph}).

Loss of a central qubit $i$ has a different effect that is illustrated in the right part of Fig.~\ref{fig:1}(d): if the fusion between one of the leaf-nodes of qubit $i$ and one of the leaf nodes of qubit $j$ succeeds, qubit $j$ gets connected to qubit $i$ and it thus needs to be measured in the $Z$-basis if qubit $i$ is lost. In the case of fusion failure, $j$ is not connected to the lost qubit $i$ and nothing needs to be done. It is assumed in this illustration that none of the fusion photons used for connecting qubits $i, j$ are lost since in this case both $i,j$ would need to be measured in the $Z$-basis anyway.

The leaf qubits are measured in the fusion process such that losses are heralded. The central qubits, on the other hand, constitute the final graph state which needs to be measured in a specific basis to perform quantum computation~\cite{Raussendorf2001}. Since the locations of the losses are typically only known when a measurement fails to detect a photon, it is unclear whether the generated graph states can be used for quantum computation in this setting~\cite{Morley2017, Lobl2023b}. In this case, the efficiency $\eta$ being above the corresponding percolation threshold $\lambda_{\eta}$ thus constitutes a necessary yet not sufficient condition for generating a graph state that is a useful resource for quantum computing. Nevertheless, we consider the possibility of all-photonic resource states as such approaches have been considered also in previous work~\cite{GimenoSegovia2015, Pant2019}. We refer to the described percolation model with all-photonic resource states as percolation model 2 for which an example is given in Fig.~\ref{fig:1}(e).

A promising alternative is using resource states where the central qubit is the spin of a quantum emitter~\cite{Lobl2023b} and we shall refer to the corresponding model as percolation model 2' which is illustrated in Fig.~\ref{fig:1}(f). In this case, the central qubits are not subject to loss, making the generated graph state suitable for measurement-based quantum computing provided that it is sufficiently large. This is the case once the efficiency $\eta$ is above the corresponding percolation threshold $\lambda_{\eta}$ of a sufficiently large fusion network. For percolation model 2', $\eta>\lambda_{\eta}$ thus is a necessary and sufficient condition for creating a graph state that is useful for quantum computing.

\subsection{Percolation model 3: adaptive fusion network}
The success probability $p_s=0.5$ of the fusions can be a limitation for the previously discussed scheme. For an implementation based on photon emitters, there is, however, a strategy to remedy this issue using adaptive fusions. We consider a scheme where the central nodes are spins in quantum emitters and fusions can be repeated on demand. To establish a connection between two emitters, each of them creates a photon entangled with it, e.g. by using a suitable optical pulse~\cite{Lindner2009, Tiurev2021}. When the fusion succeeds, we proceed with no further actions. When it fails, it can be repeated until the fusion either succeeds, a maximum number $n_{max}$ of fusion attempts is reached, or photon loss is detected. In Fig.~\ref{fig:1}(g), the case of $n_{max}=2$ is illustrated in the absence of photon loss. When there are no losses, repeating failed fusions effectively leads to a boosting of the fusion success probability above $p_s=0.5$. For the case of $n_{max}$ in Fig.~\ref{fig:1}(g), for instance, a link between two central nodes would be generated with probability $p_s+(1-p_s)\cdot p_s=0.75$ which is well above the bond percolation threshold of the fusion lattice (square lattice with $\lambda_{bond}=0.5$~\cite{Kesten1980}). We note that there are also all-optical schemes to boost the fusion success probability~\cite{Grice2011,Ewert2014,Bayerbach2023}, but the proposed repeat-until-success method is less sensitive to photon loss as additional fusion photons are only generated on demand.

As before the percolation model becomes more involved when photon loss is present in which case a sequence of fusion attempts has three possible outcomes: success, failure, or photon loss. In the success case, no photon loss occurs and the final fusion succeeds and thus creates a connection between two central nodes. Fusion failure (no connection) happens when all $n_{max}$ fusion attempts fail with no photon loss. When a photon loss is detected in any of the fusion attempts, the two quantum emitters are measured in the $Z$-basis removing them from the graph. The combination of these possibilities of fusion outcomes represents percolation model 3 as illustrated in Fig.~\ref{fig:1}(h).

The described adaptive approach is related to well-known schemes for creating spin-spin entanglement via Bell-measurements and corresponding repeat-until success schemes~\cite{Barrett2005,Lim2005,Lim2006}. An advantage of the considered scheme is that fusions do not influence other fusions (they are locally adaptive) which enables doing all fusions in parallel. That reduces the overhead compared to adaptive divide and conquer approaches~\cite{Barrett2005, Duan2005, Lim2006}. Furthermore, our scheme can be applied to emitters with only a single spin per node (in contrast to e.g. Ref.~\cite{Choi2019}).

\subsection{Lattice renormalization}
\label{sec:lattice_renormalize}
With the models described above, a large connected cluster state that scales linearly with the size of the fusion network or the initial graph state is created if the photon efficiency is above the percolation threshold $\lambda_{\eta}$. The resulting graph state does not represent a fault-tolerant lattice which eventually is required for quantum computing~\cite{Raussendorf2006}. The obtained cluster state can, however, be renormalized into such a lattice by suitable single-qubit measurements~\cite{Morley2017, Herr2018}. A complete investigation of this process is beyond the scope of this article. We therefore restrict our analysis to noting, that once the efficiency is above the threshold, scalable quantum computation is in principle possible, although different geometries may be more advantageous than others. Note further that it is required to know all imperfections (losses and fusion failures) before applying the measurement pattern performing the lattice renormalization~\cite{Morley2017, Hein2006}. This is fulfilled for percolation model 3 and percolation model 2' with a central qubit that is a spin (since fusion failure and loss of fusion photons is heralded). In these cases, an efficiency $\eta=1-p_{loss}$ above the percolation threshold $\lambda_{\eta}$ is thus a sufficient condition for generating the desired lattice. In percolation models 1 and 2, losses can be unheralded, and $\eta>\lambda_{\eta}$ thus represents only a necessary condition as discussed before.

\section{Union-find algorithm for simulating photon loss}
The percolation models introduced above give a guideline for the loss tolerance of fusion networks or graph states. In this section, we give efficient algorithms for simulating the corresponding percolation threshold $\lambda_{\eta}$. For all percolation models, it is not obvious which lattice is the best choice~\cite{Lobl2023b}. In classical bond- or site-percolation, a larger vertex degree (coordination number) of the lattice is typically helpful to lower the percolation threshold. In contrast, when a loss is compensated by removing all neighbors of a lost photon by $Z$-basis measurements (see Fig.~\ref{fig:1}(b)), a graph or fusion network with a high vertex degree is particularly sensitive to photon loss. However, a too-low vertex degree makes the graph state or fusion network fragile due to the higher bond-percolation threshold~\cite{Kieling2007, Pant2019}. To find the optimum in between and simulate various fusion lattices, a fast algorithm for computing percolation thresholds $\lambda_{\eta}$ is required. This algorithm should ideally scale linearly with the number of qubits/photons and ideally should be independent of the number $n_{\eta}$ of sampled values for $\eta$. In the following, we first describe the well-known Newman-Ziff algorithm~\cite{NewmanZiff2001} which applies to bond- or site-percolation. In the subsequent subsections, we describe modified algorithms that can be applied to the percolation models 1 to 3~\footnote{Similar modifications of the Newman-Ziff algorithm have been developed for other classical percolation models such as bootstrap or diffusion percolation~\cite{choi2019_percol}}.

The easiest way to perform classical bond- or site-percolation simulations is in the canonical ensemble where the bond (site) probabilities $p_{\text{bond}}$ ($p_{\text{site}}$) are fixed and bonds/sites are occupied randomly according to these probabilities. Connected components, their size, or whether they percolate the lattice can then be obtained with a time complexity that is linear in the system size by depth- or breadth-first graph traversal~\cite{Cormen2022}. Constructing a lattice with constant vertex degree (coordination number $z$), a number of nodes $|V|$, and a number of edges $|E|=\frac{z}{2}\cdot|V|$ takes $\orderof\left(|V|\right)$ and traversing the graph  (e.g. breadth first) takes $\orderof\left(|V|\right)$ as well~\cite{NewmanZiff2001}. In some two-dimensional percolation simulations, this scaling can be slightly improved by parsing only the boundary of a spanning cluster~\cite{Ziff1984,NewmanZiff2001}. However, this method does not solve the general issue described in the following.

\begin{figure*}[t]
\includegraphics[width=2.0\columnwidth]{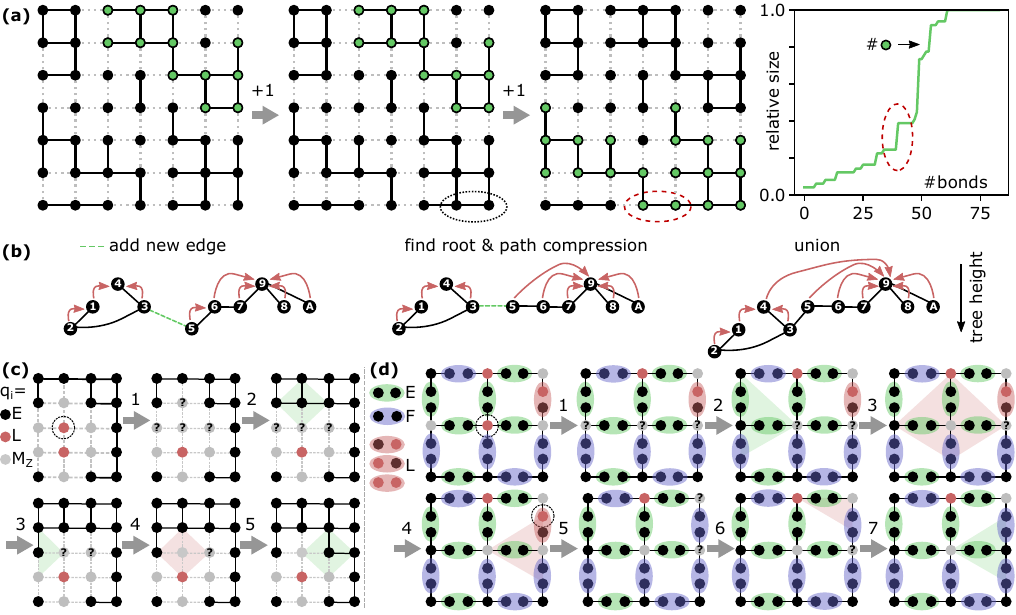}
\caption{\label{fig:2}Algorithms to speed-up percolation simulations. \textbf{(a)} The key idea of the Newman-Ziff algorithm~\cite{NewmanZiff2001} illustrated for bond-percolation. Bonds are added one by one in random order and the largest connected component of the graph (green nodes) is monitored. When two large clusters merge, there is a rapid increase in the size of the largest connected component as shown on the right. \textbf{(b)} Updating the connected graph components in classical bond-percolation~\cite{NewmanZiff2001}. In the first step (find), the root nodes of the two components that the new edge connects are identified by following the tree data structure (red arrows). In the second step (union), the two components are merged by connecting the root node of the smaller component to the root node of the larger one. \textbf{(c)} Illustration of Algorithm~\ref{alg:1} for simulating the effect of loss on a graph state (percolation model 1). Initially, two qubits are lost (red nodes) and the qubits in their neighborhood are therefore measured in the $Z$-basis (gray). When adding one missing qubit in step 1, several qubits in the neighborhood potentially need to be updated. This update is done in steps $2-5$ where the neighborhoods of all potentially affected nodes are checked for lost qubits. When none of a qubit's neighbors are lost, there is no longer any need to measure it in the $Z$-basis and it is added to the graph. \textbf{(d)} Illustration of Algorithms~\ref{alg:2} and ~\ref{alg:2b} for simulating the effect of loss when creating a large graph state by fusing star-shaped resource states (percolation models 2, 2'). Algorithm~\ref{alg:2}: in step 1, a lost central qubit is added (encircled node). All qubits that are not lost and connected to the added qubit by a successful fusion potentially need to be updated. The update is done in steps $2-4$ by applying the rules illustrated in Fig.~\ref{fig:1} to all potentially affected qubits. Algorithm~\ref{alg:2b}: after step 4 is completed, we assume in step 5 that a previously lost fusion photon is added (encircled node). The subsequent update algorithm (steps 6, 7) is similar to before with the difference that only the central node belonging to the added fusion photon and the central node belonging to its fusion partner need to be updated.}
\end{figure*}

To make a percolation simulation for $n$ different probabilities of bond occupation ($p_{\text{bond}}$) or site occupation ($p_{\text{site}}$), the computation has to be repeated $n$ times and the running time becomes $\orderof\left(n\cdot|V|\right)$~\cite{NewmanZiff2001}. When many different probabilities $p_{\text{bond}}$ or $p_{\text{site}}$ are considered, the factor $n$ becomes a major hurdle. For classical bond- and site-percolation simulations, an algorithm that avoids this issue by considering the problem in the microcanonical ensemble has been developed by Newman and Ziff~\cite{NewmanZiff2001}. In this algorithm, the number of occupied bonds/sites is fixed (not the bond/site probability as before). In the corresponding site-percolation simulations, all nodes are initially removed and added to the graph in random order one by one. (Bond percolation can be performed analogously by adding bonds as illustrated in Fig.~\ref{fig:2}(a).) Connected graph components of existing nodes are stored and when a new node is added, the corresponding data structure storing the connected components is updated. At the same time, one keeps track of the size of the largest connected component $S(i)$ and a list of booleans $B(i)$ that specifies if percolation (cluster spanning~\cite{NewmanZiff2001}) has been achieved after adding node number $i$. Having computed $\langle S(i)\rangle$ by averaging several simulations of $S(i)$, the expectation value of the largest component size can be computed as a function of $p_{\text{site}}$ by~\footnote{For most practical cases, the computation of $S(i)$ is the most time-consuming part of the simulation, although calculating Eq.~\eqref{expect} has technically a worse scaling behaviour. For large graphs however, it is only necessary to take $\orderof\left(\log|V|\right)$ elements of the sum because of the rapid decrease of the binomial distribution away from its maximum. So the running time for doing this summation for $n$ different values of $p_{\text{site}}$ becomes $\orderof\left(n\cdot\log|V|\right)$.}
\begin{equation} \label{expect}
	\langle S(p_{\text{site}})\rangle=\sum_{i=0}^{|V|}\langle S(i)\rangle\cdot\binom{|V|}{i} \cdot p_{\text{site}}^i\cdot(1-p_{\text{site}})^{|V|-i}
\end{equation}
where the binomial coefficients are computed in a normalized way using the iterative method from Ref.~\cite{NewmanZiff2001} (for large $|V|$ a Gaussian approximation of the binomial distribution \cite{Malarz2022} is an alternative). Replacing $S(i)$ by $B(i)$ in the above equation computes the probability of percolation for a given probability $p_{\text{site}}$. It has been shown that the above percolation simulation has a time-complexity of only $\orderof\left(|V|\right)$ (with no factor $n$) if a suitable union-find algorithm (see below) is used for merging graph components~\cite{Tarjan1975,NewmanZiff2001}.

Critical to implementing such an algorithm is a data structure that efficiently keeps track of the graph and its connected components as well as a method to update it when adding an edge/node. As in Ref.~\cite{NewmanZiff2001} we use a tree data structure~\cite{Galler1964} where every node is either a root node representing an isolated graph component or it points to a parent node that is part of the same graph component. An example of such a data structure and how to update it for bond-percolation is shown in Fig.~\ref{fig:2}(b) where the simulated graph is drawn in black and the tree data structure is illustrated by the red arrows. Adding a previously missing edge is done by updating the graph and the tree data structure using the union-find algorithm from Ref.~\cite{NewmanZiff2001}: following the tree data structure, the algorithm first determines the root nodes of the two graph components that the new edge connects (find). (To reduce the length of subsequently traversed paths, we employ path compression~\cite{Tarjan1975,NewmanZiff2001} such that the parent node for all nodes along the path is updated to the root node.) If the two root nodes are different, the graph components are merged by making one of the root nodes point to the other (union). To improve the performance, one attaches the smaller component to the larger one (weighted union~\cite{Tarjan1975,NewmanZiff2001}). This union operation is done in $\orderof\left(1\right)$ steps as opposed to an approach based on a look-up table storing the graph component for every node in an array. The latter would need $\orderof\left(1\right)$ for a single find operation but up to $\orderof\left(|V|\right)$ to determine all nodes of the cluster that must be updated in a union. The overall time complexity would become $\orderof\left(|V|^2\right)$ in the worst case. In contrast, the combination of path compression and weighted union using the tree data structure results in a practically linear time scaling, $\orderof\left(|V|\right)$ ~\footnote{There is a non-constant prefactor which, however, is not expected to exceed 3 in a system of any realistic size (see Refs.~\cite{Tarjan1975,NewmanZiff2001} for the discussion of the related Ackermann's function)}. This result is important for efficiently performing classical bond- or site-percolation simulations~\cite{NewmanZiff2001} and also is applied in union-find decoders for quantum error correction \cite{Delfosse2021}.

Adding a previously missing node in a site percolation simulation is done similarly~\cite{NewmanZiff2001}: the algorithm first adds the new node as an isolated single-node graph component. Then it adds the edge to the first neighbor and updates the graph like in the case of bond percolation~\footnote{Note that some edges might be labeled as missing due to a finite bond probability (or for some of the following algorithms: a finite fusion probability). In this case, only existing edges are considered.}. This procedure is repeated for all the other neighbors. We refer to this algorithm for adding a previously missing node as Algorithm~\ref{alg:0} (see Ref.~\cite{NewmanZiff2001} for more details).
\renewcommand{\thealgocf}{0}
\begin{algorithm}
\caption{add new node}\label{alg:0}
Add a new node to the graph and update the tree data structure with the algorithm from Ref.~\cite{NewmanZiff2001}
\end{algorithm}

In the following, we describe algorithms that use the tools described so far to efficiently simulate the percolation models 1 to 3 and obtain the corresponding percolation threshold $\lambda_{\eta}$. Algorithm~\ref{alg:1} applies to the loss tolerance of a given cluster state (percolation model 1). Algorithms~\ref{alg:2},~\ref{alg:2b} apply to fusion networks (percolation model 2). Algorithm~\ref{alg:2} applies to lost central qubits of the star-shaped resource states and thus only is used when the central qubits are photons that can be lost. Algorithm~\ref{alg:2b} applies to fusion photons and thus is used independently of whether the central node is a photon (percolation model 2) or a quantum emitter (percolation model 2'). Finally, Algorithm~\ref{alg:3} applies to a repeat-until-success fusion network (percolation model 3).

\subsection{Update rule for percolation model 1}
First, we consider percolation model 1 where an existing photonic cluster state suffers loss. Similar to the Newman-Ziff algorithm~\cite{NewmanZiff2001} we assume initially that all photons are lost and then add them one by one keeping track of the largest connected graph component or checking if a graph component percolates. However, new rules are required to update the graph when a previously lost photon is added. Updating the cluster state after adding a previously lost qubit $i$ is complicated by the fact that the neighborhood $N(i)$ of a lost qubit used to be measured in the $Z$-basis~\cite{GimenoSegovia2015}. When adding a missing qubit, those neighbors potentially have to be updated and added to the graph if there is no longer any need to measure them in the $Z$-basis. We assign a label $q_i$ to every central qubit of a resource state: $q_i=L$ meaning that the qubit is lost, $q_i=M_Z$ meaning that the qubit needs to be measured in the $Z$-basis, or $q_i=E$ meaning that the qubit exists (not lost nor has to be measured in the $Z$-basis, and therefore is part of the graph/cluster state). Algorithm~\ref{alg:1} describes the corresponding data structure update when changing the label of a qubit in a cluster state from lost to not lost (decreasing the number of lost photons by one).

\renewcommand{\thealgocf}{1}
\begin{algorithm}
\caption{add lost qubit ($q_i=L$) to the graph state}\label{alg:1}
\tcc{step 1: change label of lost qubit}
$q_i:=M_Z$\;
\tcc{step 2: list potentially affected qubits}
$\Tilde{V} := \{j \in N(i) \mid q_j=M_Z\} \cup \{i\}$\;
\tcc{step 3: update and add qubits}
\For{$j \in\ \Tilde{V}$}{
    \If{$\left(\nexists\ k \in N(j): q_k=L\right)$}{
        $q_j:=E$\;
        add node $j$ via Algorithm~\ref{alg:0}\;
    }
}
\end{algorithm}

The algorithm is illustrated in Fig.~\ref{fig:2}(c). In the first step, the label of the lost qubit $i$ is changed to $q_i=M_Z$ indicating that it is not lost anymore. In the second step, a list $\Tilde{V}$ of all potentially affected qubits is created (all qubits in the neighborhood $N(i)$ of qubit $i$ as well as qubit $i$ itself). In the final step, the neighborhood of all qubits in $\Tilde{V}$ is checked for lost qubits. If a lost qubit is still found in the neighborhood of a qubit $j\in\Tilde{V}$, no update needs to be done. If no lost qubit is found in the neighborhood $N(j)$, the label of the qubit is updated to $E$. In the latter case, qubit $j$ was only affected by the loss of qubit $i$ and can be added to the graph once qubit $i$ is not lost anymore. Adding qubits to the graph is done with Algorithm~\ref{alg:0} from Ref.~\cite{NewmanZiff2001}.

\subsection{Update rule for percolation models 2 and 2'}
Experimentally more realistic than percolation model 1 is constructing a large graph state by fusing many small star-shaped resource states (percolation model 2)~\cite{GimenoSegovia2015,Lobl2023b}. In this percolation model, fusions between \textit{leaf} nodes are used to establish connections and every photon (either in the center of a star-shaped resource state or a fusion photon on the \textit{leaf}) can be lost with a uniform probability $p_{loss}=1-\eta$. We assign different labels to every edge $e_{ij}$ ($i$, $j$ are the numbers of the central qubits of the star-shaped resource states that the edge connects upon successful fusion): $e_{ij}=E$ when the corresponding fusion was successful, $e_{ij}=F$ when the fusion failed but no photon was lost, and $e_{ij}=L$ when at least one of the photons on the edge is lost. In the case of loss, we also need to keep track of whether one ($c_{ij}=1$) or no ($c_{ij}=0$) fusion photon on edge $e_{ij}$ is present (not lost), i.e. whether both or just one photon was lost.

When adding a previously lost photon, there are two cases: the photon can be a central qubit of the star-shaped resource state or it can be a fusion photon on a leaf node. We first consider the case of adding a lost central qubit to the corresponding fusion network (applies to percolation model 2, not 2'). The graph update is done by Algorithm~\ref{alg:2} (see Fig.~\ref{fig:2}(d) for an illustration).
\renewcommand{\thealgocf}{2a}
\begin{algorithm}
\caption{add lost central qubit ($q_i=L$) to the fusion network}\label{alg:2}
\tcc{step 1: change label of lost qubit}
$q_i:=M_Z$\;
\tcc{step 2: list potentially affected qubits}
$\Tilde{V} := \{j \in N(i)\mid q_j=M_Z \land \ e_{ij}=E\} \cup \{i\}$\;
\tcc{step 3: update and add qubits}
\For{$j \in \Tilde{V}$}{
    \If{$\left(\nexists\ k \in N(j): q_k=L \land\ e_{jk}=E\right) \land$ \\$\left(\nexists\ k \in N(j): e_{jk}=L\right)$}{
        $q_j:=E$\;
        add node $j$ via Algorithm~\ref{alg:0}\;
    }
}
\end{algorithm}
Here the neighborhood $N(i)$ refers to all qubits in the center of other star-shaped resource states that one tries to connect to qubit $i$ by a fusion. In the first step, the label of the lost qubit is changed to $q_i:=M_Z$ as in Algorithm~\ref{alg:1}. In the second step, a list $\Tilde{V}$ of all potentially affected qubits is created. A qubit $j\neq i$ can only be affected by a change of qubit $i$ when there is a connection between both qubits obtained by a successful fusion ($e_{ij}=E$). In the final step, all potentially affected qubits $j \in \Tilde{V}$ are updated. A qubit can only be added ($q_j:=E$) when it is not connected to a lost qubit by a successful fusion ($\nexists\ k \in\ N(j): q_k=L \land\ e_{jk}=E$) and no fusion photon on any of its edges is lost ($\nexists\ k \in\ N(j): e_{jk}=L$).

Next, we consider the case where the lost photon is a fusion photon between two central qubits $i$, $j$. This case applies to both percolation model 2 and percolation model 2' with an example shown in the lower part of Fig.~\ref{fig:2}(d). Here, Algorithm~\ref{alg:2} needs to be slightly modified: first, the edge $e_{ij}$ must be updated if and only if the fusion partner of the new qubit is not lost ($c_{ij}=1$ before the update), otherwise $e_{ij}=L$ remains unchanged and nothing else needs to be done. In the former case, the edge label is updated to $e_{ij}=F$ or $e_{ij}=E$ with probabilities of $1-p_s=p_s=1/2$, and steps 2 and 3 are performed. Second, the list of affected central qubits in step 2 must be changed to $\Tilde{V}=\{i, j\}$. Algorithm~\ref{alg:2b} describes the corresponding graph update rule.

\renewcommand{\thealgocf}{2b}
\begin{algorithm}
\caption{add lost fusion photon on edge $e_{ij}$ of fusion network}\label{alg:2b}
\tcc{step 1: check if any update is needed}
$c_{ij}:=c_{ij}+1$\;
\uIf{$c_{ij}=2$}{
    with probability $p_s$: $e_{ij}:=E$\;
    otherwise: $e_{ij}:=F$\;
} \Else {
    skip steps 2, 3\;
}
\tcc{step 2: list potentially affected qubits}
$\Tilde{V} := \{k \in \{i, j\}\mid q_k \neq L\}$\;
\tcc{step 3: update and add qubits}
\For{$l \in \Tilde{V}$}{
    \If{$\left(\nexists\ k \in N(l): q_k=L \land\ e_{lk}=E\right) \land$\\ \hspace{1.2cm} $\left(\nexists\ k \in N(l): e_{lk}=L\right)$}{
        $q_l:=E$\;
        add node $l$ via Algorithm~\ref{alg:0}\;
    }
}
\end{algorithm}

\subsection{Update rule for percolation model 3}
\begin{figure}[t]
\includegraphics[width=1.0\columnwidth]{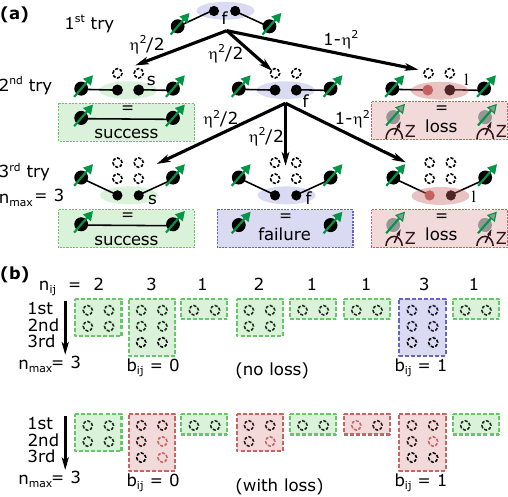}
\caption{\label{fig:adaptive}\textbf{(a)} Adaptive scheme where fusions are repeated until either success, heralded photon loss, or until $n_{max}=3$ trials are reached. We have assumed here that the first fusion attempt fails. The scheme can be realized using quantum emitters with a spin (drawn as black dots with green arrows) and proceeds by adjacent quantum emitters generating new fusion photons if an attempt to entangle them with a fusion fails. When one or more fusion photons are lost, both quantum emitter spins need to be reinitialized, or measured in the $Z$-basis. For every single fusion attempt, the corresponding outcome is indicated by the the color of the ellipses that indicate the fusion as well as a letter: \textit{s} and green ellipse for fusion success (probability $\eta^2p_s=\eta^2/2$), \textit{f} and blue ellipse for fusion failure (probability $\eta^2(1-p_s)=\eta^2/2$), \textit{l} and red ellipse for a fusion with a loss (probability $1-\eta^2$). The dashed rectangular boxes indicate the final outcome of the repeat-until success fusion when following the corresponding path. \textbf{(b)} Schematic illustration of the data structure used for keeping track of the fusion results when sweeping the number of lost photons. Every dashed box represents a fusion that is repeated in case it does not succeed. The color represents the final outcome of the fusion (green: success, blue: failure, red: loss) and every dashed circle inside the boxes represents a fusion photon with red color indicating that the photon is lost. In the absence of loss (upper part), the overall fusion succeeds except when all individual fusion attempts fail ($b_{ij}=1$). As long as there is one lost photon in the box representing the overall sequence of fusion attempts ($c_{ij}<2\cdot n_{ij}$), the fusion terminates by re-initializing the spin qubit.}
\end{figure}
So far, we have explored how the Newman-Ziff algorithm can be modified to simulate the effect of photon loss in fusion networks where every fusion is performed once, independent of the outcome of other fusions. In percolation model 3, the outcome of previous fusions determines whether a fusion is repeated or not and the algorithms developed so far thus cannot be applied. In Algorithm~\ref{alg:2}, for instance, we initially labeled all photons as lost ($L$) and added them one by one. In percolation model 3, the overall number of photons is not constant and it is therefore unclear how many photons exist and where in the graph they are used. We show how a modification of the Newman-Ziff algorithm nevertheless can be applied to the adaptive repeat-until-success fusion network described by percolation model 3.

In percolation model 3, a fusion is repeated until it either succeeds, fails after a maximum number of repetitions $n_{max}$, or heralds photon loss before completion. An exemplary tree diagram for the different fusion outcomes and the associated probabilities is given in Fig.~\ref{fig:adaptive}(a). The probabilities that the fusion attempts on a certain edge succeed ($p_{\Tilde{S}}$), fail ($p_{\Tilde{F}}$), or terminate with a loss ($p_{\Tilde{L}}$) are~\cite{Lim2006, Gliniasty2023}:
\begin{align}
    &p_{\Tilde{F}} = \left(\eta^2(1-p_s)\right)^{n_{max}} = \alpha^{n_{max}}\label{eq:pF}\\
    &p_{\Tilde{S}} = \eta^2\cdot p_s\cdot\sum_{k=0}^{n_{max}-1}(\eta^2(1-p_s))^k = \alpha\cdot\frac{\alpha^{n_{max}}-1}{\alpha-1}\label{eq:pS}\\
    &p_{\Tilde{L}} = 1 - p_{\Tilde{F}} - p_{\Tilde{S}} \label{eq:pL}
\end{align}
where we have used $p_s=0.5$ for the success rate of a single fusion attempt and defined $\alpha\equiv\eta^2/2$. In the case of photon loss, the two quantum emitter qubits are removed from the graph by $Z$-basis measurements as before~\footnote{This step is not required for systems where entangler and memory qubit are two different systems that can be coupled/decoupled on demand~\cite{Choi2019,Pompili2021,Simmons2023}. For quantum emitters like quantum dots, such techniques are, however, only at a very early stage~\cite{Appel2024}.}

Our algorithm to simulate percolation model 3 consists of two main steps. In the first step, we assume that there is no photon loss and determine for every edge $e_{ij}$ how many fusion attempts are required for the fusion to succeed. We label this number of fusion attempts $n_{ij}$ and it is determined by tossing a random coin until the coin indicates success or we reach the maximum number of allowed fusions $n_{max}$. The probability distribution for the frequency of $n_{ij}$ thus decays exponentially for large values of $n_{ij}$. When none of the $n_{max}$ fusions succeed, we keep track of this by a Boolean array $b_{ij}$ where $b_{ij}=0$ indicates fusion success and $b_{ij}=1$ fusion failure. The number $n_{ij}$ determines how many potentially lost photons we have to consider for the edge $e_{ij}$, namely exactly $2\cdot n_{ij}$ as illustrated in Fig.~\ref{fig:adaptive}(b). The reason is that after $n_{ij}$ fusion attempts, no further fusions on $e_{ij}$ are performed because the fusion either succeeded, or failed when $n_{ij}=n_{max}\land b_{ij}=1$, or because there has been a heralded photon loss before. So without loss, the overall number of fusion photons of all fusion attempts is $\sum_{ij}2\cdot n_{ij}$.

In the second step, the fusion photons of all the edges are labeled as lost and are added one by one in random order while updating the connected graph components. In that process, an edge keeps the label $e_{ij}=L$ until all its $2\cdot n_{ij}$ fusion photons are not lost anymore. We count for every edge $e_{ij}$ the number of photons that are not lost with a counter variable $c_{ij}$ (initially zero and increasing when a fusion photon is added on edge $e_{ij}$). A change of the graph needs to be done once $c_{ij}$ reaches $2\cdot n_{ij}$ (before that, the attempt to establish a connection between nodes $i$ and $j$ always terminates with a photon loss)\footnote{This algorithm works because the following is fulfilled: assume a fraction $\Tilde{\eta}=\sum_{ij}c_{ij} / \sum_{ij}2\cdot n_{ij}$ of all fusion photons has been updated to not lost. Then the probabilities that the result of the fusion attempts on a random edge $e_{ij}$ is labeled as failure, success, or loss are equal to the actual probabilities $p_{\Tilde{F}}, p_{\Tilde{S}}, p_{\Tilde{F}}$ in Eq.~\eqref{eq:pF},~\ref{eq:pS},~\ref{eq:pL}. Thus Eq.~\eqref{expect} (with $|V|$ being replaced by the number of all photons $\sum_{ij}2\cdot n_{ij}$ and the averaging being done outside the sum) yields the expectation value of the largest graph component size as a function of $\Tilde{\eta}$.}. Algorithm~\ref{alg:3} describes how this update is done. As the central qubits $q_i$ are quantum emitters (and thus cannot be lost), Algorithm~\ref{alg:3} differs from Algorithm~\ref{alg:2b} in steps 2, 3).

\renewcommand{\thealgocf}{3}
\begin{algorithm}
\caption{add previously lost fusion photon on edge $e_{ij}$ of adaptive fusion network.}\label{alg:3}
\tcc{step 1: check if any update is needed}
$c_{ij}:=c_{ij}+1$\;
\uIf{$c_{ij}=2\cdot n_{ij}$}{
    \uIf{$b_{ij}=1$}{
        $e_{ij}=F$\;
    } \Else{
        $e_{ij}=E$\;
    }
} \Else {
    skip all following steps\;
}
\tcc{step 2: list potentially affected qubits}
$\Tilde{V} := \{i, j\}$\;
\tcc{step 3: update and add qubits}
\For{$l \in \Tilde{V}$}{
    \If{$\left(\nexists\ k \in N(l): e_{kl}=L\right)$}{
        $q_l:=E$\;
        add node $l$ via Algorithm~\ref{alg:0}\;
    }
}
\end{algorithm}

Finally, we note that an alternative method for boosting the fusion success probability uses ancilla entangled photon pairs~\cite{Grice2011} and this approach could be modeled similarly. The probabilities for fusion success and loss of a fusion photon in this case differ from Eqs.~\eqref{eq:pS},~\eqref{eq:pL} as $2^n$ photons are used for every fusion ($n=1$ being the standard fusion) to achieve a success probability of $1-1/2^n$. We therefore would have: $p_{\Tilde{F}} = \eta^{(2^n)}/2^n$, $p_{\Tilde{S}} = \eta^{(2^n)}\cdot(1-1/2^{n})$, $p_{\Tilde{L}} = 1 - \eta^{(2^n)}$. Using $2^n$ photons per fusion can be simulated with the approach from before by setting $2\cdot n_{ij}=2^n$. Once all fusion photons exist, the fusion is set to failure with a probability of $1/2^{n}$ and success otherwise. Alternatively, since the number of photons is fixed for the all-photonic boosting, it would be possible to make a modified Newman-Ziff algorithm by adding lost fusions one by one instead of virtually adding all individual lost photons. The size of the largest connected graph component is then obtained as a function of the number of fusions with no lost photons. Using the relation between $\eta$ and $p_{\Tilde{L}}$ one can convert such a simulation into the largest connected graph component as a function of $\eta$ by rescaling the $x$-axis. For percolation model 3, such a simplification is, however, not possible because $p_{\Tilde{S}}/p_{\Tilde{F}}$ is not constant as a function of $\eta$. Even if one would compute $p_{\Tilde{S}}/p_{\Tilde{F}}$ (by converting $p_{\Tilde{L}}$ into $\eta$) in every step where a previously lossy fusion is added, all the fusions that were added before would appear with a ratio $p_{\Tilde{S}}/p_{\Tilde{F}}$ that is inconsistent with the current fraction of non-lossy fusions (resp. $p_{\Tilde{L}}$). Also when fusion photons as well as other photons are lost (Algorithm~\ref{alg:2}), the only feasible approach is virtually adding photons one by one and doing the required data structure update in every step.

\section{Results and running time}
\begin{figure*}
\includegraphics[width=2.0\columnwidth]{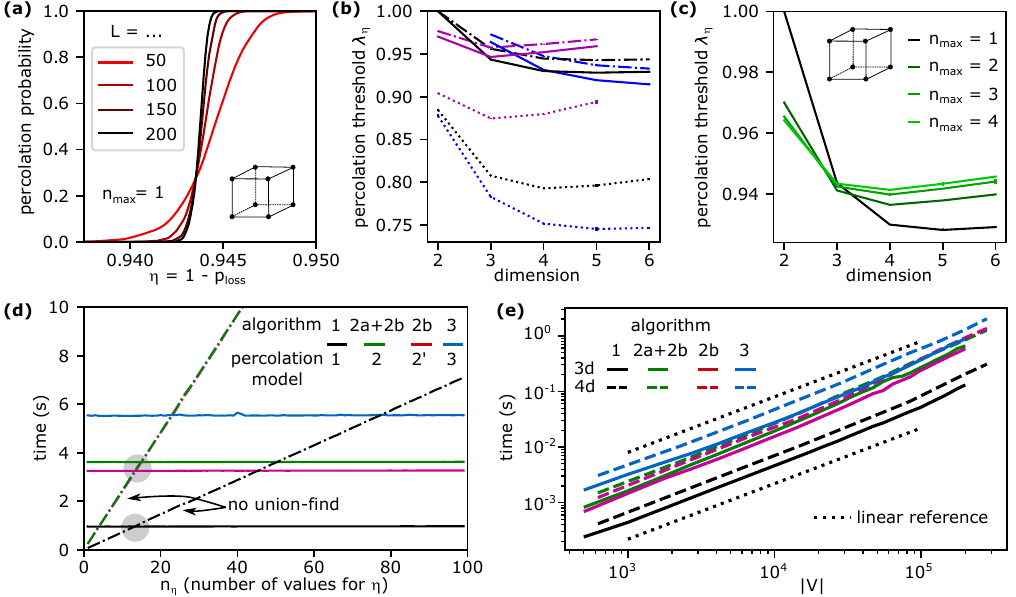}
\caption{\label{fig:results} \textbf{(a)} Simulation of the simple cubic fusion lattice for percolation model 2' (equivalent to percolation model 3 with $n_{max}=1$). The size of the simulated lattices is specified by the parameter $L$ which is the number of unit cells along all three dimensions. \textbf{(b)} Percolation threshold $\lambda_{\eta}$ for the diamond (blue), hypercubic (black), and bcc+hc lattices (purple). The dotted lines show the results for percolation model 1, the dashed lines are simulations of all-photonic fusion networks (percolation model 2), and the solid lines correspond to simulations with quantum emitters as the central qubits (percolation model 2'). \textbf{(c)} Percolation threshold of the hypercubic fusion lattice for different values of $n_{max}$ (percolation model 3). \textbf{(d)} Running time for simulating a 3D simple-cubic lattice with $L=100$ and so $L^3=10^6$ central nodes, averaged over $20$ repetitions~\footnote{Used system: Rocky Linux 8.7 (Green Obsidian), AMD Opteron(TM) Processor 6276 (single core used per simulation)}. The simulations are performed for $n_{\eta}$ different values of $\eta=1-p_{loss}$. When using Algorithms~\ref{alg:1},~\ref{alg:2},~\ref{alg:2b},~\ref{alg:3}, the simulation time is constant, independent of $n_{\eta}$ (solid lines). Without these algorithms, the simulation time increases linearly with $n_{\eta}$ (dashed lines). The gray circles indicate the number $n_{\eta}$ where the modified Newman-Ziff algorithms become more efficient. (For percolation model 3, we have only implemented a modification of the Newman-Ziff algorithm and thus no dashed line is shown.) \textbf{(e)} Running time of different percolation simulations as a function of the lattice size $|V|$. In the case of fusion networks (percolation models 2, 2', 3), $|V|$ represents the number of central qubits, not counting the fusion photons. The running time of a full percolation simulation (averaged over $300$ repetitions) scales linearly with $|V|$ for a three-dimensional (solid lines) and a four-dimensional (dashed lines) simple cubic lattice. The dotted lines indicate linear scaling for reference.}
\end{figure*}
We have implemented the described algorithms for the various percolation models and the source code can be found in Ref.~\cite{Lobl2023}. More details on the source code and how we verify parts of the implementation can also be found in Appendix~\ref{section_code}.

In this section, we use these implementations for several percolation simulations. Fig.~\ref{fig:results}(a) shows an exemplary percolation simulation (percolation model 2') of the simple cubic lattice for different lattice sizes. From such a simulation, we can obtain the corresponding percolation threshold $\lambda_{\eta}$. Fig.~\ref{fig:results}(b) shows the obtained percolation thresholds for the percolation models 1, 2, and 2' on several different lattices. More simulation results of percolation models 2, 2' can be found in Ref.~\cite{Lobl2023b}.

Figs.~\ref{fig:results}(c) shows the percolation thresholds for percolation model 3 using the hypercubic (hc) fusion lattices~\cite{VanderMarck1998}. These simulations are performed for different values of the maximum number of fusion repetitions ($n_{max}$). We observe that the adaptive scheme investigated here improves the loss tolerance for the low-dimensional lattices but makes it worse in higher dimensions. This trend is also observed for other fusion lattices that we investigated. The reason for the worse loss tolerance is that repeating the fusions until success increases the chance that the fusion terminates by photon loss. Using all-photonic boosting~\cite{Grice2011, Ewert2014} would have a similar issue. Of course, the illustrated adaptive scheme is not ideal: fusions are for instance performed even if it is known that one of the involved quantum-emitter qubits must be measured in the $Z$-basis due to a loss in a previous fusion. Problems like this can be avoided with more sophisticated adaptive schemes.

Next, we consider the running times of the percolation simulations. When using Algorithms~\ref{alg:1}, ~\ref{alg:2},~\ref{alg:2b}, or ~\ref{alg:3}, we expect that the running time is independent of the number $n_{\eta}$ of simulated photon loss probabilities. Without these algorithms, a new graph traversal needs to be performed for every efficiency $\eta=1-p_{loss}$ leading to a time that is linear in $n_{\eta}$. Fig.~\ref{fig:results}(d) shows the measured running times as a function of $n_{\eta}$ confirming the expected time scalings with $n_{\eta}$. We find that the simulations using the modified Newman-Ziff algorithms become more efficient when about a dozen different values for $\eta$ are simulated. The speed advantage is key for simulating the loss tolerance of large fusion networks~\cite{Lobl2023b}.

Furthermore, we expect that the running time as a function of the system size (number of vertices $|V|$ in the cluster state or fusion network) scales identically to Ref.~\cite{NewmanZiff2001} meaning practically linear with $|V|$. The reason is that the overhead of Algorithms~\ref{alg:1} and~\ref{alg:2} compared to Algorithm~\ref{alg:0} is a constant prefactor that depends on the vertex degree of the simulated lattice. This intuition is supported by Fig.~\ref{fig:results}(e) where we find an approximately linear scaling of the running time when increasing the size of the simulated lattice.

\section{Summary and Outlook}
We have shown that it is possible to use a modification of Newman and Ziff's algorithm~\cite{NewmanZiff2001} for non-standard percolation models motivated by photon loss in cluster states and fusion networks. We have considered ballistic fusion networks where many resource states are connected by simultaneous probabilistic fusions and an adaptive scheme where fusions are repeated until success. The developed algorithms speed up the simulation of the considered architectures which enables running optimizations over many fusion lattices~\cite{Lobl2023b}. Mitigating the photon loss with such optimizations is highly important as loss is arguably the biggest challenge for photonic quantum computing and networking.

We have considered star-shaped resource states but our algorithms also apply when resource states with a graph structure of a caterpillar tree are used (like star-shaped resource states these states can be generated by a single quantum emitter~\cite{Paesani2022}). The only condition for our algorithms to work is that all fusion photons are \textit{leaf} nodes of the resource state graphs and thus only have a single neighbor which also may apply to some graph states that cannot be generated by a single quantum emitter~\cite{Economou2010, Li2022}). Whether similar algorithms can be applied when fusing other types of states such as linear chain graph states~\cite{Lindner2009, Thomas2022} could be investigated in future works. Furthermore, policies other than repeat-until-success could be used in the adaptive scheme to decide whether a fusion is applied or not. It would be interesting to investigate if similar modifications of the Newman-Ziff algorithm could be applied to such percolation models.

Finally, we hope that our results might be inspiring or useful for other applications in quantum information: percolation models can be applied to quantum networks~\cite{Acin2007}, the bond-percolation threshold of the syndrome graph determines the loss threshold of various topological quantum error-correction codes~\cite{Stace2009, Barrett2010, Auger2018, Stricker2020, Bartolucci2021}, and qubit loss in color codes can be simulated with modified percolation models~\cite{Vodola2018, Amaro2020}. In addition, logical errors in topological codes are operators that span around the code~\cite{Dennis2002}. The transition where errors plus the decoder output give a percolating operator is thus a lower bound for the error threshold~\cite{Hastings2014}. All these points indicate that similar algorithms as the ones developed here could improve the analysis of quantum error correction in general.

\section{Acknowledgment}
We would like to thank Love A. Pettersson for fruitful discussions and Daniel L\"{o}bl for support with the C-implementation. We are grateful for financial support from Danmarks Grundforskningsfond (DNRF 139, Hy-Q Center for Hybrid Quantum Networks). S.P. acknowledges funding from the Marie Skłodowska-Curie Fellowship project QSun (nr. 101063763), from the VILLUM FONDEN research grant VIL50326, and support from the NNF Quantum Computing Programme.

\appendix
\renewcommand\thefigure{\thesection.\arabic{figure}} 

\section{source code}
\label{section_code}
\subsection{General description}
Our simulation program is written in the programming language C and is provided as an open-source repository~\cite{Lobl2023}. It can be used to simulate $\lambda_{\eta}$ for percolation models 1, 2, 2', and 3 with various lattice geometries in several dimensions. Furthermore, it enables simulating standard bond- or site-percolation using the Newmna-Ziff algorithm. It also applies to site-bond or bond-site~\cite{Wattendorff2024} percolation simulations (where one parameter being either the bond- or site-probability is fixed and the other parameter is simulated over a continuous range using the Newman-Ziff algorithm). The following explanation focuses on percolation models 2 and 2' but the other models are implemented very similarly.

Every percolation simulation consists of three main steps: In the first step, a certain lattice with a fixed size is constructed. We use an adjacency list to represent the graphs/fusion lattices keeping the data structure analogous to Ref.~\cite{NewmanZiff2001}. Therefore, other lattices can be easily built or taken from the literature where a similar implementation is used~\cite{Malarz2022}. In the second step, the effect of the random process of photon loss and fusion failure is simulated. Since the fusion probability is a fixed parameter given by the physical implementation, we first randomly set fusions to success/fail. The effect of photon loss is then simulated using the described algorithms where all photons are labeled as lost initially and then added one by one. During that process, we keep track of the largest component size $S(i)$ or a Boolean array $B(i)$ specifying if percolation (cluster spanning) has been achieved, with $i$ specifying the number of photons that are not lost. This second part of the simulation is repeated several times for averaging. We estimate the percolation threshold for a given lattice size as the fraction of non-lost photons $\eta$ where cluster spanning is achieved on average when adding photons one by one~\footnote{Several different methods could be used instead to determine the percolation thershold~\cite{Ziff2010, Bastas2014, Malarz2022_b}}. In the third step, we determine the percolation threshold in the limit of infinite lattice size by compensating for finite-size effects with an extrapolation (see Ref.~\cite{VanderMarck1998}). Furthermore, to obtain the expectation value for the probability of percolation or the largest connected component size as a function of the parameter $\eta$, we use Eq.~\eqref{expect}. In this way, a plot like, for instance, the one shown in Fig.~\ref{fig:results}(a) is obtained.

For the graph construction in the first step, we provide several functions that construct lattices such as hypercubic, diamond, bcc, and fcc in a dimension of choice. In particular, our implementation allows simulating all lattices from Ref.~\cite{Kurzawski2012} including their higher-dimensional generalizations, all lattices from Ref.~\cite{VanderMarck1998} except the Kagome lattice, and hypercubic lattices with extended neighborhoods~\cite{Malarz2005, Xun2020, Zhao2022}.

\subsection{Verification of the implementation}
To verify our lattice constructions, we have performed classical site-percolation simulations of lattices with known percolation thresholds~\cite{VanderMarck1998, Kurzawski2012}. These simulations can be found in Ref.~\cite{Lobl2023b} where we have simulated two- to six-dimensional lattices that, for periodic boundaries, correspond to various $k$-regular graphs (meaning that every node has exactly $k$ neighbors) with $k\in[3..60]$. Additionally, we compute here a few bond-percolation thresholds. Note that none of these simulations represent a central part of this work yet an agreement with literature values is a useful consistency check to verify that the lattices are defined and implemented correctly.

For the bond-percolation thresholds of the generalized diamond lattices~\cite{VanderMarck1998}, we find $\lambda_{bond}=0.6529(10),\,0.3893(7),\,0.2709(18),\,0.2072(15),\,0.1646(43)$ for dimensions $2-6$ respectively which is in agreement with Refs.~\cite{Sykes1964, Xu2014, VanderMarck1998}. Furthermore, we find $\lambda_{bond}=0.2491 \pm 0.0002$ for the bond-percolation threshold of the simple-cubic lattice (roughly agrees with Ref.~\cite{Wang2013}) which also represents the syndrome graph of the RHG lattice~\cite{Raussendorf2006}. For the latter, we find a bond-percolation threshold of $\lambda_{bond}=0.3845(1)$ and a site-percolation threshold of $\lambda_{site}=0.4220(5)$.

The agreement of several simulations with literature values verifies that our implementation of the corresponding lattices is correct. However, the studied percolation models for photon loss as well as algorithms~\ref{alg:1},~\ref{alg:2},~\ref{alg:2b},~\ref{alg:3} have not been considered in the literature before. The bond- and site-percolation simulations, therefore, do not say anything about the correctness nor the correct implementation of these algorithms. To verify that the implementation of our algorithms is correct, we have implemented several redundant functions for the corresponding percolation models without any Newman-Ziff-type algorithm (see readme file of the code repository~\cite{Lobl2023} for more information). We have performed several tests for which the results were consistent between the different implementations.

\bibliography{lit}

\end{document}